\documentclass[aps,pra,twocolumn]{revtex4}   
\newcommand {\be}{\begin{equation}}
\newcommand {\ee}{\end{equation}}
\newcommand {\bey}{\begin{eqnarray}}
\newcommand {\eey}{\end{eqnarray}}
\usepackage{epsfig}
\usepackage{amsfonts}
\usepackage{mathbbol}

\begin{document}

\title{Exponential complexity and ontological theories of quantum mechanics}

\author{A. Montina}
\affiliation{Dipartimento di Fisica, Universit\`a di Firenze,
Via Sansone 1, 50019 Sesto Fiorentino (FI), Italy}

\date{\today}

\begin{abstract}
Ontological theories of quantum mechanics describe a single system 
by means of well-defined classical variables and attribute the quantum 
uncertainties to our ignorance about the underlying reality 
represented by these variables. We consider the general class of ontological
theories describing a quantum system by a set of variables with Markovian 
(either deterministic or stochastic) evolution. 
We provide the first proof that the number of continuous variables 
can not be smaller than $2N-2$, $N$ being the Hilbert space dimension.  
Thus, any ontological Markovian theory of quantum mechanics requires a 
number of variables which grows exponentially with the physical 
size. This result is relevant also in the framework of quantum Monte 
Carlo methods.
\end{abstract}
\maketitle
\section{introduction}
In classical mechanics, the number of variables describing the state of a particle
ensemble scales as the number of particles, thus the calculation time and memory 
resources which are necessary in a numerical simulation are generally a polynomial
function of the physical size. This
friendly property of classical systems seems to be lost at the microscopic
level, when quantum effects become relevant.
It is well-known that the definition of a quantum state requires
a number of resources growing exponentially with the number
of particles. This characteristic is at the basis of the exponential
complexity of quantum mechanics ($QM$) and forbids at present to efficiently
simulate the dynamics of many-body systems, unless some approximation
is involved or particular problems are 
considered~\cite{nummet,trahan,oriols,montina_dice}.

The origin of the exponential growth can be understood with the
following example. A quantum scalar particle is associated with a complex
wave-function $\psi(\vec x)$ which lives in the three-dimensional space
and evolves in accordance with the Schr\"odinger equation. Discretizing
each coordinate with $R$ points and evaluating the spatial derivatives
by finite differences, the Schr\"odinger equation is then reduced to
$2 R^3$ real ordinary differential equations. One could interpret $\psi$
as a physical field which pilots the particle and intuitively conclude
that $M$ interacting particles are trivially described by $M$ complex
wave-functions in the three-dimensional space. Thus, the number
of variables would grow linearly with the number of particles.
Instead, in the standard description of quantum phenomena, an ensemble
of $M$ particles is associated with a single complex wave-function 
$\psi(\vec x_1,...,\vec x_M)$ which lives in the configuration
space of the whole system. If each coordinate is discretized with
$R$ points, the number of real variables we have to integrate grows
exponentially with $M$ and is given by $2 R^{3 M}$.

In general terms, if two quantum systems $A_1$ and $A_2$ are associated
with the Hilbert spaces ${\cal H}_1$ and ${\cal H}_2$, then the
composite system $A_1+A_2$ is associated with the tensorial product
${\cal H}_1\otimes{\cal H}_2$. This characteristic was at the basis
of the Born's argumentations against the realistic interpretation of
the wave-function by Schr\"odinger, which would imply
an exponential growth of physical variables. In the Born's interpretation
the wave-function does not represent real quantities, but it
is merely a mathematical tool of the theory enabling to evaluate
the probabilities of physical events. It is similar to the classical
concept of probability distribution, that has the form 
$\rho(\vec x_1,\vec p_1,...,\vec x_M,\vec p_M)$ for an ensemble
of $M$ particles and lives in the phase space of the whole system.
In spite of this similarity, the Born's interpretation constitutes
a deep departure from classical statistical theories, where each
system is described by well-defined (ontological) quantities and the 
probability distributions merely provide a statistical 
(epistemological) description of them (in the following, we use
the terms "classical", "ontological" and "realistic" as synonyms).

This departure is
not unavoidable, indeed ontological theories of quantum phenomena exist, 
such as Bohm mechanics~\cite{bohm,bell,Durr},
which describes single systems by means of well-defined variables
(ontic variables) determining the outgoing values of measurements.
However the Born's
argumentation constitutes the main difficulty of these approaches,
since they promote the abstract wave-function to the rank of
a classical real field. Thus, it is natural to ask whether there
exists an ontological theory of quantum phenomena which uses an
alternative representation of a single system and is not subject to
the exponential growth of the number of variables. This is not only
a foundational problem, but it has also a concrete relevance,
since such a theory would provide a new revolutionary quantum 
Monte Carlo method ($QMC$) for many-body 
problems~\cite{oriols,montina_dice,montina}.
Questions pertaining to the computational complexity of
$QMC$ are discussed in Ref.~\cite{troyer}.

In this article, we consider this problem from a general point of view 
and give the first proof that the exponential growth of the number of
variables is a common feature of any ontological Markovian (deterministic
or probabilistic) theory of quantum systems. More precisely we prove 
that, for a system with a finite $N$-dimensional Hilbert space,
the number of continuous ontic variables can not be smaller than
$2 N-2$.
As a consequence, the Born's criticism does not concern only the
pilot-wave theories, but every realistic Markovian theory. If we
assume $QM$ exact, then this result suggests two
choices, either to reject ontological causal theories of quantum
systems or to accept the exponential growth of the ontic variables
as an intrinsic feature of quantum phenomena. 
After this work was completed, we became aware that similar
questions were discussed by L. Hardy~\cite{hardy}. He proved
that the number of ontic states can not be finite, but did not placed 
constraints on the dimensionality of the ontic state space.
As far as we know, the question on the exponential growth of the
ontological dimensionality was posed for the first time in 
Refs.\cite{montina,montina_dice}. 
Very recently, similar studies has been tackled in Ref.~\cite{harrigan}. 
However, until now no answer to this open question was given.

One could object that the use of the multi-particle wave-function
as a variable field in ontological models is an obvious requirement,
because of Bell's theorem on entangled states. Thus, our proof
would not be necessary. However Bell's theorem imposes only a
non-locality condition on realistic theories and says nothing
about the dimensionality of the ontic state space.
One can not reject {\it a priori} the possibility of a non-local
realistic theory with a space of variables smaller than the
Hilbert space without an explicit proof. 

In section~\ref{sect_class} we give the motivations for this study and 
introduce the general properties that an ontological theory of $QM$ 
has to satisfy. In section~\ref{sect_examples} we provide some examples 
and in section~\ref{dim_redu}
discuss a particular case of dimensional reduction of
the ontic state space. In
section~\ref{sect_proof} the proof of the theorem on the classical
space dimension is presented and its consequences for quantum
Monte Carlo methods are sketched. Finally, the conclusions are drawn in 
the last section.

\section{Classical theories of quantum mechanics}
\label{sect_class}
The state of a quantum system is described by a trace-one density 
operator $\hat\rho$ which is Hermitian and non-negative defined, i.e.,
it satisfies the properties
\be\label{prop_ma_dens}
\begin{array}{l}
{\it Tr}\hat\rho=1 \\
\hat\rho^\dagger=\hat\rho \\
\langle\phi|\hat\rho|\phi\rangle\ge0\text{  for every }|\phi\rangle.
\end{array}
\ee
When the maximal information on the system is obtained, the
quantum state is described by a Hilbert space vector $|\psi\rangle$
and the density operator is the projector $|\psi\rangle\langle\psi|$.
A von Neumann measurement is associated with a Hermitian operator 
$\hat M=\sum_k m_k|k\rangle\langle k|$,
where $m_k$ are the possible results and the vectors $|k\rangle$ are 
orthonormal. If $\hat M$ is not degenerate, when $\hat M$ is measured
and the system is in the pure state $|\psi\rangle$, the probability
of obtaining result $m_k$ is $p_k=|\langle k|\psi\rangle|^2$. 
If $m_k$ is obtained, the quantum state is projected into
$|k\rangle$. This is the general framework of $QM$.

In spite of the evident success of $QM$ in explaining
secular experimental results, there are at least two reasons
to ask for an alternative reformulation of the theory.
The first one concerns the ambiguity in the state projection 
rule, which requires one to mark a boundary between the fuzzy 
microscopic quantum world and the macroscopic well-defined
observations. This ambiguity has at present no practical
consequence, since the quantum predictions are practically 
insensitive to the boundary, once the quantum domain is
taken sufficiently large. This insensitivity is related to
decoherence phenomena~\cite{zurek}. Our work is motivated
by another more practical reason. In classical mechanics,
the number of variables which specify the state scales
linearly with the physical size, thus a numerical
simulation of dynamics is generally a polynomial complexity
problem. Conversely, the definition of a quantum state
requires an exponentially growing number of resources,
making the numerical integration of the Schr\"odinger
equation impossible even for a small number of particles.
Many approximate methods are used in order to
circumvent this problem, such as quantum Monte Carlo and
semi-classical methods, the Hartree-Fock approximation, 
the density-functional theory and so on. However, at present
no general numerical method is known which is able to solve
in polynomial time quantum many-body dynamics. 
It is interesting to observe that in the standard interpretation 
the wave-function does not represent a physical field, but it 
provides a complete statistical information on an infinite ensemble of
realizations, thus the exponential
growth of resources in the solution of the Schr\"odinger equation
is not surprising. Indeed, this occurs also in classical mechanics with 
the definition of the multi-particle probability distribution
$\rho(\vec x_1,\vec p_1,...,\vec x_M,\vec p_M)$. For example, 
consider the problem of $M$ mutually coupled classical Brownian particles, 
which are described by the stochastic equations 
\be\label{stocha}
\frac{d{\vec v}_i}{dt}=\sum_{j\ne i}{\vec F}_{ij}({\vec x}_i,{\vec x}_j)
-\gamma {\vec v}_i+{\vec \eta}_i(t),
\ee
where ${\vec x}_i$, ${\vec v}_i$, ${\vec F}_{ij}$, $\gamma$ and 
${\vec \eta}_i\equiv(\eta_i^1,\eta_i^2,\eta_i^3)$ are the spatial 
coordinates, the velocity, the interaction force, the coefficient of viscosity 
and the noise term, respectively. The two-time correlation function
of the noise term is
$\langle\eta_i^k(t)\eta_j^l(t')\rangle=g\delta_{ij}\delta_{kl}\delta(t-t')$.
The masses are set equal to $1$.
The solution of these equations is the trajectory
of a single realization and its evaluation is a polynomial problem. Equation~(\ref{stocha}) 
is associated with the following Fokker-Planck equation
\be\label{fokker}
\frac{\partial\rho}{\partial t}=\sum_i\left[\frac{\partial}{\partial {\vec v}_i}\cdot
(\gamma{\vec v}_i-\sum_{j\ne i}{\vec F}_{ij})+\frac{g}{2}
\frac{\partial^2}{\partial{\vec v}_i^2}-\frac{\partial}{\partial{\vec x}_i}
\cdot{\vec v}_i\right]\rho.
\ee
Its direct numerical integration is an exponential complexity
problem. The difference of complexity between
Eq.~(\ref{stocha}) and Eq.~(\ref{fokker}) is not amazing, since
the first equation describes the trajectory of a single realization,
conversely the second one provides a complete statistical description of 
an infinite number of realizations. Since this complete information
is in general out of the experimental domain, a detailed evolution of the
 multi-particle probability distribution is not required
and it is practically sufficient to evaluate the averages of some
quantities over a finite number of trajectories by means of a Monte
Carlo method. In $QM$, a similar approach is used 
for thermal equilibrium problems and, with some approximations, in 
dynamical problems (quantum Monte Carlo methods).
The point of this discussion is the following.
Suppose, in accordance with the Einstein's view, that
there exists a more fundamental theory which does not provide a statistical 
description on ensembles and characterizes each quantum system by means of a 
set of well-defined physical variables. The dynamical laws of this ontological 
theory and the Schr\"odinger equation would be similar
to Eqs.~(\ref{stocha},\ref{fokker}) of our example, respectively.
Thus, it is natural to pose the following question: is the trajectory
simulation in this fundamental theory a polynomial problem? more
precisely, does the ontological space dimension grow polynomially with the
physical size? In this case, the fundamental theory would provide
in a natural way a revolutionary quantum Monte Carlo method.
This is the non-obvious core question of this article and we will find
that the answer is negative, i.e., the exponential complexity
is not related to the ensemble description, but it is a general feature of 
any ontological Markovian theory of quantum phenomena.
In the following subsections, we introduce the general properties of 
such theories.
\subsection{Kinematics}
\label{kine}
We characterize a single system 
by means of a set of continuous and discrete ontological variables, say
$x_1$,$x_2$,...,$x_w$ and $s_1$,...,$s_p$, respectively. In the following
we use synthetically the symbol $X$ for this set, i.e., 
$X=\{x_1,...x_w,s_1,...s_p\}$. For a single system, it takes a 
well-defined value $X(t)$ at any time $t$.
When a quantum system is
prepared in a state $|\psi\rangle$, the ontological variable
takes a value $X$ with a probability $\rho(X)$ which depends on 
$|\psi\rangle$. 
$\rho(X)$ has to satisfy the following conditions
\bey\label{norm_cond}
\int dX\rho(X)=1 \\
\label{pos_cond}
\rho(X)\ge0.
\eey
In principle, it is possible that $|\psi\rangle$ does not fix unequivocally
$\rho$; 
this one could then depend on the specific experimental setup used to 
prepare the pure state. Furthermore in the quantum formalism the pure 
state preparation deletes the memory of the previous history. This is
not necessarily true in the ontological theory. In order not to lose in 
generality, we assume that each pure state can be associated with different
probability distributions which can depend on the specific experimental setup 
and the previous history. This prevents us from writing a single-valued 
functional relation $|\psi\rangle\rightarrow\rho(X|\psi)$. Let $\cal C$ be
a set which contains at least one element, we write the mapping
\be\label{mapping}
|\psi\rangle\rightarrow\{\rho(X|\psi,\eta),\eta\in{\cal C}\}.
\ee
This over-labeling of $\rho$ can be found also in the positive 
$P$-functions~\cite{gardiner}, mainly used in quantum optics and degenerate 
boson gases, and in Ref.~\cite{spekkens}, where $\eta\in\cal C$ is
identified as context for the quantum state preparation.
Obviously, if $|\psi\rangle\langle\psi|\ne|\psi'\rangle\langle\psi'|$, 
then $\rho(X|\psi,\eta)\ne\rho(X|\psi',\eta')$ for any $\eta$ and $\eta'\in{\cal C}$. 
Thus, Function~(\ref{mapping}) can be inverted and we have
\be\label{map2}
|\psi\rangle\langle\psi|={\hat D}(\rho),
\ee
where the operator $\hat D(\rho)$ is a function whose domain is the
image of the functional $\rho(X|\psi,\eta)$. 

Equation~(\ref{mapping}) and its equivalent form, Eq.~(\ref{map2}),
are our first hypothesis.
As a practical example, consider a single mode of the electromagnetic
field, whose annihilation and creation operators are $\hat a$ and
$\hat a^\dagger$, respectively. The coherent state $|\alpha\rangle$ of 
the mode is 
\be
|\alpha\rangle=e^{-|\alpha|^2/2}\sum_{n=0}^\infty\frac{(\alpha\hat a^\dagger)^n}{n!}
|0\rangle,
\ee
where $|0\rangle$ is the vacuum state and $\alpha$ is a complex number. 
By means of the coherent state, it is possible to define some
quasi-probability distributions associated with the quantum states. The 
Glauber distribution $P_G(\alpha)$ of the state $|\psi\rangle$ is defined as follows,
\be
\int d\alpha P_G(\alpha)|\alpha\rangle\langle\alpha|\equiv|\psi\rangle\langle\psi|.
\ee
A Glauber distribution exists for any quantum state,
but in general it is highly singular and non-positive and can not be interpreted
as a probability distribution [Eq.~(\ref{pos_cond})]. The positive-$P$ distribution
$P(\alpha,\beta)$ is a generalization of $P_G$ and is such that
\be\label{pos-P}
\int d\alpha d\beta P(\alpha,\beta)\hat B(\alpha,\beta)
\equiv|\psi\rangle\langle\psi|,
\ee
where $\hat B(\alpha,\beta)\equiv e^{(|\alpha|^2+|\beta|^2-2\beta^*\alpha)/2}|\alpha\rangle\langle\beta|$.
It is possible to prove that each quantum state $|\psi\rangle$ is associated with 
a positive-$P$ distribution~\cite{gardiner}. Equation~(\ref{pos-P}) is a concrete
example of Eq.~(\ref{map2}), $X$ being the variable set $\{\alpha,\beta\}$.
Note that a single mode has an infinite dimensional Hilbert space, conversely
the $\{\alpha,\beta\}$ space is four-dimensional, i.e, we can represent a quantum 
state as a statistical ensemble on a reduced space of variables. These variables and 
$P(\alpha,\beta)$ are analogous to the variables $\{\vec x_i,\vec v_i\}$ 
in Eq.~(\ref{stocha}) and the probability distribution $\rho$ in Eq.~(\ref{fokker}), 
respectively. The dimensional reduction program would look, at this stage, feasible.
However, we have still to define the dynamics of $X$ and the connection between
$X$ and the measurements. 

\subsection{Dynamics}
\label{subsect_dynamics}
In quantum mechanics, the state evolution of a conservative system from
time $t_0$ to $t_1$ is described by a unitary operator $\hat U(t_1;t_0)$,
i.e., we have the Markovian and deterministic
evolution
\be\label{unit}
|\psi(t_1)\rangle=\hat U(t_1;t_0)|\psi(t_0)\rangle.
\ee
We retain in the ontological theory the Markovian property and define
a conditional probability $P(X,t_1|\bar X,t_0)$ such that
$\rho_1(X)=\int d\bar X P(X,t_1|\bar X,t_0)\rho_0(\bar X)$,
where $\rho_0(X)$ and $\rho_1(X)$ are two probability distributions
associated with $|\psi(t_0)\rangle$ and $|\psi(t_1)\rangle$, respectively.

We assume that every unitary evolution is physically attainable. This
hypothesis rests on the fact that, in quantum computers, every unitary
evolution is in principle feasible by means of a finite number of
quantum gates, that have a physical implementation (see for example Chapter 
4 of Ref.~\cite{nielsen}). As a consequence, every unitary operator has
to be associated with a conditional probability. As for the probability
distribution, in general each unitary operator can be mapped to 
many different conditional probabilities which can depend on the
the physical implementation of the evolution. Thus, we introduce
a set $\cal E$ which contains at least an element and define the
mapping 
\be
\hat U\rightarrow\{P(X|\bar X,\hat U,\chi),\chi\in{\cal E}\}.
\ee
The label $\chi\in{\cal E}$ identifies the context for the
unitary evolution~\cite{spekkens}.

The conditional probabilities have to satisfy the following conditions:
\begin{enumerate}
\item For any $\hat U$ and $\chi\in{\cal E}$
\bey
\label{cond_pos}
P(X|\bar X,\hat U,\chi)\ge0, \\
\label{cond_norm}
\int d\bar X P(\bar X|X,\hat U,\chi)=1.
\eey
\item For any $\hat U$, $\chi\in{\cal E}$, $|\psi\rangle$ and $\eta\in{\cal C}$, there exists a 
$\eta_1\in{\cal C}$ such that
\be\label{eq_evol}
\rho(X|\hat U\psi,\eta_1)=\int d\bar X P(X|\bar X,\hat U,\chi)\rho(\bar X|\psi,\eta);
\ee
\item For any $\hat U_{1,2}$ and $\chi_{1,2}\in \cal E$, there exists a $\chi_3\in{\cal E}$ 
such that
\be\begin{array}{c}
\label{Prod_Evo}
\int dX_i P(\bar X|X_i,\hat U_2,\chi_2) P(X_i|X,\hat U_1,\chi_1)=  \\
P(\bar X|X,\hat U_2\hat U_1,\chi_3).
\end{array}
\ee
\end{enumerate}
Note that the integral symbol $\int dX$ synthetically indicates 
the integration and sum over the continuous and discrete variables.

The set $\{P(\bar X|X,\hat U,\chi),\chi\in{\cal E}\}$ is 
an equivalence class labeled by $\hat U$. The set of all the equivalence
classes is a group, whose identity is
$\{P(\bar X|X,\mathbb{1},\chi),\chi\in{\cal E}\}$.

These are the general hypotheses for the dynamics in the ontological theory. 
In the case of the positive-$P$ distribution, it is possible to define conditional
probabilities which satisfy these conditions, but we will not show it. 

It is important for our purposes to deduce some properties of $P(X|\bar X,\hat U,\chi)$
and $\rho(X|\psi,\eta)$. We introduce the following

{\bf Definition} $\bf 1$. We denote by $I(\psi,\eta)$ the support of the
probability distribution $\rho(X|\psi,\eta)$, i.e., the smallest closed set with 
probability $1$ and define $I(\psi)$ as follows
\be
I(\psi)=\{I(\psi,\eta),\eta\in C\}
\ee
In practice, if the set of values $X$ is countable, then 
$X\in I(\psi,\eta)\Leftrightarrow\rho(X|\psi,\eta)\ne0$. For continuous spaces
this is still true apart from a zero probability set. In order
not to be pedantic, we assume that these negligible sets are null, i.e., we
assume that
our probability distributions have the same properties of discrete distributions.

{\bf Property} {\bf 1}. If $\bar X\in I(\psi)$, then the support of 
$P(X|\bar X,\hat U,\chi)$ is a subset of $I(\hat U\psi)$. Equivalently,
if $\bar X\in I(\psi)$ and $P(X|\bar X,\hat U,\chi)\ne0$, then
$X\in I(U\psi)$.

If the $X$ space is discrete, this property is a consequence of 
Eqs~(\ref{pos_cond},\ref{cond_pos},\ref{eq_evol}). 
The proof is as follows: if $\bar X\in I(\psi)$ then there exists an $\eta$ such that
$\rho(\bar X|\psi,\eta)\ne0$. Thus, if $X$ is an element of the support of 
$P(X|\bar X,\hat U,\chi)$, 
then exists an $\eta_1$ (second enumerated conditions) such that
$\rho(X|\hat U\psi,\eta_1)=\sum_{\tilde X}P(X|\tilde X,\hat U,\chi)\rho(\tilde X|\psi,\eta)
\ge P(X|\bar X,\hat U,\chi)\rho(\bar X|\psi,\eta)\ne0$,
i.e., $X$ is an element of $I(\hat U\psi)$. For continuous spaces, Property~$1$ is
true apart from unimportant sets with zero probability. As previously said,
we will assume them null.

\begin{figure}[h!]
\epsfig{figure=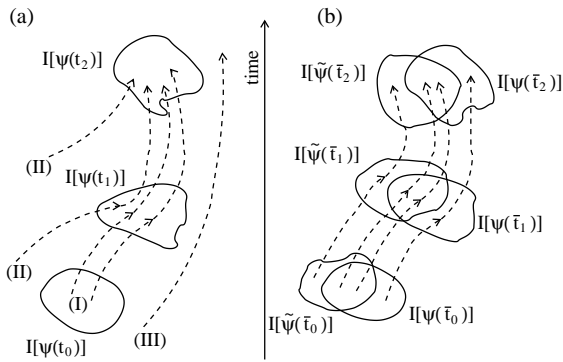,width=7.5cm}
\caption{Visual representation of the evolution of $I[\psi(t)]$ in the ontic state
space. (a) Classical states in $I[\psi(t_0)]$ at time $t_0$ [labeled with (I)] 
can not escape from  $I[\psi(t)]$ at subsequent time $t>t_0$. However, some states outside
$I[\psi(t_0)]$ [labeled with (II)] can jump into $I[\psi(t)]$. Other states (III) may
exist which remain always outside $I[\psi(t)]$. (b) The evolution of the support $I$
for two different states $|\psi\rangle$ and $|\tilde\psi\rangle$. After a suitable
transient evolution,
no trajectory can escape from or to jump into $I[\psi(t)]$ and $I[\tilde\psi(t)]$.}
\label{fig1}
\end{figure}

Property~$1$ can formulated in these terms. If $X(t)$ is the deterministic/stochastic
trajectory of the ontological variable and $X(t_0)\in I(\psi)$ at time $t_0$, then at
a subsequent time $t_1>t_0$ $X(t_1)\in I[\hat U(t_1;t_0)\psi]$. In general, 
this property is not invariant for time inversion and $X(t_0)\notin I(\psi)$ 
does not implies that $X(t_1)\notin I[\hat U(t_1;t_0)\psi]$. The non-invariance
is due to the fact that a backward Markovian process is not in general a
Markovian process. Thus, the set $I[\psi(t)]$
is as a black hole, the trajectories $X(t)$ can jump into it but can not escape
from it. This is illustrated in Fig.~\ref{fig1}a. The points (I), (II) and (III) in
the figure are defined in the caption.

It is reasonable to assume that this confluence of the trajectories into
$I[\hat U(t_1;t_0)\psi]$ corresponds to a transient behavior which becomes negligible
after a suitable series of unitary evolutions. This is equivalent to say
that for any $|\psi'\rangle$, there exists a probability distribution
$\rho(X|\psi',\eta)$ whose support contains only points of type (I) and (III)
(Fig.~\ref{fig1}b).

In the following, we assume that, after a suitable transient evolution, the variable 
$X$ evolves towards a classical subspace where the time symmetry is fulfilled
and consider only this subspace as the space of ontic states.
Thus, we enunciate

{\bf Property 2}. For any $|\psi\rangle$, $\hat U$ and $\chi\in{\cal E}$, if 
$\bar X\notin I(\psi)$ and $P(X|\bar X,\hat U,\chi)\ne0$, then $X\notin I(U\psi)$. 

Equivalently, if $X(t)$ is a deterministic/stochastic trajectory
and $X(t_0)\notin I(\psi)$ at time $t_0$, then at a subsequent time 
$t_1>t_0$ $X(t_1)\notin I[\hat U(t_1;t_0)\psi]$ (Fig.~\ref{fig1}b).

It is interesting to note that the actual state of a system we seek to
describe in the laboratory might not have undergone the suitable transient
evolution and not be in a region with the time symmetry property. For example,
this might occur for a system sufficiently isolated since the early universe.
However, it important to realize that the ontic state space is fixed once and for
all and it must be able to describe every potential future evolution of the
system. For our purpose, it is sufficient to know that there exists a region
with the time symmetry property established by Property 2. The constraint
on the dimensionality that we will find for this sub-region will be
valid for the whole classical space. 

Properties~1-2 will be fundamental for our proof in Section~\ref{sect_proof}.

\subsection{Measurements}
\label{subsect_measure}
Let $\hat M$ be a Hermitian operator with eigenvectors $|k\rangle$ and eigenvalues $m_k$,
$k$ being an integer index. 
When a measurement of $\hat M$ is performed on the state $|\psi\rangle$, the result
$m_k$ is obtained with probability 
\be
p_k=|\langle k|\psi\rangle|^2
\ee
if $m_k$ is not degenerate.

In the ontological theory, we introduce a
conditional probability $P_{\hat M}(k|X)$ such that~\cite{montina}
\bey\label{cond_meas_1}
\sum_k P_{\hat M}(k|X)=1, \\
\label{cond_meas_2}
P_{\hat M}(k|X)\ge0,  \\
\label{cond_meas_3}
\int dX P_{\hat M}(k|X) \rho(X|\psi,\eta)=|\langle k|\psi\rangle|^2,
\eey
for any $\eta\in \cal C$ and $|\psi\rangle$. Equation~(\ref{cond_meas_3}) makes
the classical measurement rule equivalent to the Born's rule.

When $P_{\hat M}(k|X)$ takes only the values $0$ and $1$, the
ontological theory is called "dispersion-free" and the variable $X$ determines exactly
the measurement results. However, this property is not necessary for our scope and
will not be assumed. 
It is useful to note that each Hermitian operator $\hat M$ may be associated with many 
conditional probability distributions, 
i.e., the measurement results may depend on the physical implementation of
the measurement of $\hat M$. 
In order to account for this dependence, we have to
introduce a set $\cal D$, akin to $\cal C$ and $\cal E$, and let the conditional
probability depend on $\tau \in \cal D$, that is, denote the conditional
probability by $P_{\hat M}(k|X,\tau)$~\cite{spekkens}. This function has
to satisfy properties~(\ref{cond_meas_1}-\ref{cond_meas_3}) for any 
$\tau\in{\cal D}$.

For our purpose, it is sufficient to
consider only trace-one projectors and prove the following

{\bf Property 3}. If $|\psi\rangle$ and $|\psi_\perp\rangle$ are two orthogonal states,
then the supports of $\rho(X|\psi,\eta_1)$ and $\rho(X|\psi_\perp,\eta_2)$ do not
contain common elements, for every $\eta_1$ and $\eta_2$, i.e., 
$I(\psi)\cap I(\psi_\perp)=\oslash$. Proof: Assume {\it ab absurdo} the opposite and let
$X_0$ be a value such that $\rho(X_0|\psi,\eta_1)$ and $\rho(X_0|\psi_\perp,\eta_2)$
are not equal to zero.
Let $P(1|X)$ be the conditional probability of obtaining $1$ with the projective
measurement $|\psi\rangle\langle\psi|$ and a fixed $\tau$. By Eq.~(\ref{cond_meas_3}), 
we have
\be\label{first_step}
\int dX P(1|X)\rho(X|\psi,\eta_1)=1.
\ee
Since $\rho(X|\psi,\eta_1)$ is positive and normalized to one [Eqs.~(\ref{norm_cond},\ref{pos_cond})],
and $\rho(X_0|\psi,\eta_1)\ne0$, by
Eq.~(\ref{first_step}) we have that $P(1|X_0)=1$. This is obvious if the set of values $X$
is countable. As said in Section~\ref{subsect_dynamics}, this is true also in the continuous 
case apart from events with zero probability which can be neglected. Similarly, we
have
\be\label{second_step}
\int dX P(1|X)\rho(X|\psi_\perp,\eta_2)=0.
\ee
Since $\rho(X_0|\psi_\perp,\eta_2)\ne0$, we have that $P(1|X_0)=0$,
in contradiction with the previous deduction. Thus, $X_0$ can not be a
common element of $I(\psi)$ and $I(\psi_\perp)$.
In simple words, since $|\psi\rangle$ and $|\psi_\perp\rangle$ correspond
to mutually exclusive events, they can not be associated with the same value of the
ontological variable.
This property has been used in Ref.~\cite{spekkens} to derive a no-go theorem for 
noncontextual hidden variable models, and a similar property has been
used in Ref.~\cite{hardy} to derive the "ontological excess baggage theorem".

It is interesting to note that every positive distribution introduced in
quantum optics, such as the positive-$P$ and the Husimi $Q$ functions do not
provide an ontological  description of quantum mechanics, since in general their
support is the whole phase space and two orthogonal quantum states can have 
overlapping probability distributions. 

We have completed our characterization of an ontological theory of $QM$.
In the following section, we introduce some ontological models of simple quantum systems and
show that their space dimension is always equal to or larger than $2 N-2$, where $N$
is the Hilbert space dimension. 

\section{Two examples of ontological theories}
\label{sect_examples}
\subsection{two-state quantum system}
\label{sect2}
We consider the ontological model of a two-state system reported in Ref.~\cite{kochen}.
The classical variable $X$ is a three-dimensional unit vector, which we denote by $\vec v$.
Let $|-1\rangle$ and $|1\rangle$ be two orthogonal states,
we associate each quantum state $|\psi\rangle\equiv\psi_{-1}|-1\rangle+\psi_1|1\rangle$
with the following probability distribution in $X$,
\be\label{prob_2states}
\rho(\vec v|\psi)=\frac{1}{\pi}\vec v\cdot\vec w(\psi)\theta[\vec v\cdot\vec w(\psi)],
\ee
where $\theta$ is the Heaviside function and the three components of $\vec w(\psi)$ are
\be\begin{array}{l}
\label{bloch_vector}
w_1(\psi)\equiv\psi_{-1}^*\psi_1+\psi_1^*\psi_{-1},   \\
w_2(\psi)\equiv -i\psi_{-1}^*\psi_1+i\psi_1^*\psi_{-1},   \\
w_3(\psi)\equiv |\psi_{-1}|^2-|\psi_1|^2,
\end{array}
\ee
i.e., $\vec w(\psi)$ is the Bloch vector of $|\psi\rangle$. $\vec w(\psi)$ is the symmetry axis
of the probability distribution, whose support is a hemisphere (the region where $\rho$ is different
from zero).

We write the Hamiltonian as
\be
\hat H=\sum_{k=1}^3 h_k(t) \hat\sigma_k,
\ee
$\hat\sigma_k$ and $h_k(t)$ being the Pauli matrices and three real time-dependent coefficients,
respectively. The vectors $|\pm1\rangle$ are the eigenstates of $\hat\sigma_3$ with
eigenvalues $\pm1$.
It is easy to prove that the equation of motion of $\vec w(\psi)$ is
\be
\frac{d w_i}{dt}=2\sum_{jk}\epsilon_{ijk}h_j(t) w_k,
\ee
where $\epsilon_{ijk}$ is the Levi-Civita symbol. Thus, the probability distribution evolves
rigidly according to the Louville equation
\be
\frac{\partial\rho}{\partial t}=-2\sum_{ijk}\epsilon_{ijk}h_j(t) v_k\frac{\partial\rho}{\partial v_i},
\ee
which corresponds to the deterministic equation for the ontological variable $\vec v$
\be
\frac{d v_i}{dt}=2\sum_{jk}\epsilon_{ijk}h_j(t) v_k.
\ee
The measurement rule is as follows: the probability of an event associated with the vector $|\phi\rangle$ 
is
\be\label{cond_prob_spin}
P(\phi|\vec v)\equiv\theta[\vec w(\phi)\cdot\vec v].
\ee
It is possible to prove that
\be
\int d\vec v P(\phi|\vec v)\rho(\vec v|\psi)=|\langle\phi|\psi\rangle|^2,
\ee
the second side being the probability of the event $|\phi\rangle$ according to the Born rule.
Thus, we can describe a two-state system as a classical one in a space with the dimension equal to 
$2 N-2$, where $N=2$ is the Hilbert space dimension.

It is useful to remark that, in this model, each vector $\vec v$ is not associated with
only one quantum state, i.e., probability distributions associated with different quantum
states can overlap. In other words, if we define ${\cal S}(\vec v)$ as
the set of Hilbert space vectors $|\psi\rangle$ such that $\rho(\vec v|\psi)\ne0$,
then ${\cal S}(\vec v)$ contains infinite elements. Note that this model satisfies 
property~$3$, i.e., if $|\psi\rangle$ and $|\psi_\perp\rangle$ are two orthogonal vectors and
$|\psi\rangle\in{\cal S}(\vec v)$, then $|\psi_\perp\rangle\notin{\cal S}(\vec v)$
(This property will be further discussed in Section~\ref{sect_proof}). 

\subsection{Higher dimension of the Hilbert space and Kochen-Specker theorem}

Classical dispersion-free models are possible also for higher
dimensional quantum systems. Here we discuss a simple example
introduced in the first chapter of Ref.~\cite{bell}. It is
very artificial, but shows that ontological formulations of
quantum mechanics are possible. 

We consider a quantum system associated with an $N$-dimensional
Hilbert space. Let $|1\rangle$, $|2\rangle$,...,$|N\rangle$ be an
orthonormal basis. 
We associate the quantum state
\be
|\psi\rangle=\sum_k\psi_k|k\rangle
\ee
with the probability distribution
\be
\rho(\chi_1,....,\chi_N,\lambda|\psi)\equiv \prod_k \delta(\chi_k-\psi_k),
\ee
where the ontic state space is spanned by the $N$ complex variables $\chi_k$ and $\lambda$,
which takes values in the real interval $[0,1]$ with uniform probability.
Obviously, the ontological variables $\chi_k$ evolves deterministically as $\psi_k$,
i.e., by means of the Schr\"odinger equation
\be
i\hbar\frac{\partial \chi_k}{\partial t}=\sum_l \langle k|\hat H|l\rangle \chi_l.
\ee
We can consider $\lambda$ a constant of motion. At this point, we have
to write a conditional probability for events. Let $|\phi(1)\rangle$,...$|\phi(N)\rangle$
be a set of orthonormal vectors associated with events. If the projective 
operator $\hat P^{(1)}=|\phi(1)\rangle\langle\phi(1)|$ is measured, the probability
of the event $\phi(1)$ is
\be
P(1|\psi)\equiv|\langle\phi(1)|\psi\rangle|^2.
\ee
It is easy to prove that
\be\label{cond_prob_1}
P(1|\psi)=\int d^{2N}\chi \int_0^1d\lambda P(1|\chi,\lambda) \rho(\chi,\lambda|\psi),
\ee
where
\be
P(1|\chi,\lambda)\equiv \theta[|\langle\chi|\phi(1)\rangle|^2-\lambda]
\ee
with
$|\chi\rangle\equiv\sum_k\chi_k|k\rangle$. Thus, the function $P(1|\chi,\lambda)$
can be interpreted as the conditional probability of the event $1$. 
If the projective operator 
$|\phi(2)\rangle\langle\phi(2)|$ is subsequently measured, the probability of the
event $\phi(2)$ is
\be
P(2|\psi)\equiv|\langle\phi(2)|\psi\rangle|^2.
\ee
We want to find a conditional probability $P(2|\chi,\lambda)$ such that
\be
P(2|\psi)=\int d^{2N}\chi \int_0^1d\lambda P(2|\chi,\lambda) \rho(\chi,\lambda|\psi).
\ee
Since the two events $\phi(1)$ and $\phi(2)$ are mutually exclusive, $P(1|\chi,\lambda)$ and
$P(2|\chi,\lambda)$ cannot be different from zero for the same values of the conditional
variables. Bearing this in mind, we put
\be
P(2|\chi,\lambda)=\theta[|\langle\chi|\phi(1)\rangle|^2+|\langle\chi|\phi(2)\rangle|^2-\lambda]-
P(1|\chi,\lambda),
\ee
i.e., $P(2|\chi,\lambda)$ is $1$ for $|\langle\chi|\phi(1)\rangle|^2<\lambda<
|\langle\chi|\phi(1)\rangle|^2+|\langle\chi|\phi(2)\rangle|^2$ and zero elsewhere.
Analogous constructions can be made for the other projective measurements 
$|\phi(k)\rangle\langle\phi(k)|$. 

It is interesting to note that the conditional probability of the event $\phi(2)$
depends by construction on $|\phi(1)\rangle$, i.e., a different choice of the first 
projective measurement modifies the outgoing result of the second one. This 
characteristic is called {\it contextuality} and is unavoidable when the Hilbert
space dimension is higher than $2$, as established by the Kochen-Specker
theorem~\cite{kochen,mermin}.

We have shown that it is possible to construct an ontological theory which fulfills the
three conditions established in Section~\ref{sect_class}. The classical variables are
$2 N+1$ in number. However, since $\sum_k|\chi_k|^2=1$ and the global phase is unimportant,
the manifold of ontic states can be reduced to $2N-1$. Another dimension can be 
eliminated if we give up the dispersion-free property. A very simple example of model
which is not dispersion-free and satisfies our three general conditions is obtained with the 
following probability
distribution and conditional probability for the state $\psi$ and the event $\phi$,
\be
\rho(\chi_1,....,\chi_N|\psi)\equiv \prod_k \delta(\chi_k-\psi_k),
\ee
\be
P(\phi|\chi)=|\langle\phi|\chi\rangle|^2.
\ee
The corresponding ontological manifold has $2N-2$ dimensions. Although this example sounds
trivial, it shows that an ontological theory in a $2N-2$ dimensional space which satisfy
our conditions is possible. As we will prove in 
Section~\ref{sect_proof}, this dimensional value is also the lowest possible one. 

\section{Dimensional reduction of the ontological space in a particular case}
\label{dim_redu}
In this section we discuss an example of dimensional reduction of the ontological
space. This reduction is possible for particular quantum states
and measurements. We will consider a bosonic mode and
show that there exists a four-dimensional manifold in the Hilbert space 
whose elements can be represented in a two-dimensional classical space. 
Since this example is very well-known in literature,
its discussion will be brief. For more details, see for example 
Ref.~\cite{montina_dice} and references cited there in.

We consider a one bosonic mode with a Hamiltonian quadratic in the
annihilation and creation operators $\hat a$ and $\hat a^\dagger$. The results can
be extended to the case of a higher number of modes. 
The Wigner distribution of
a quantum state $|\psi\rangle$ is by definition the function
\be\label{Wigner}
W(\alpha)\equiv\frac{1}{\pi^2}\int \langle\psi|e^{\lambda\hat a^\dagger-\lambda^*\hat a}|\psi\rangle
e^{\lambda^*\alpha-\lambda\alpha^*}d^2\lambda,
\ee
where $\alpha$ is a complex number and the domain of integration is the complex plane.

Let $\hat q$ and $\hat p$ be two Hermitian operators such that $\hat a=(\hat q+i\hat p)/\sqrt2$,
they satisfy the canonical commutation relation $[\hat q,\hat p]=i$. In the basis of the
eigenvectors $|x\rangle$ of $\hat q$, Equation~(\ref{Wigner}) becomes
\be\label{Wigner2}
W(q,p)=\frac{1}{2\pi}\int_{-\infty}^\infty d x \psi(q+x/2)\psi^*(q-x/2)e^{-i x p},
\ee
where $q+ip\equiv\alpha$ and $\psi(x)\equiv\langle x|\psi\rangle$.

The Wigner function satisfies the identity
\be
\int dq dp W(q,p)=1.
\ee
In general, it can take negative values, but for particular states it
is positive in the whole phase space $(q,p)$ and can be interpreted as a probability
distribution. This is the case of the Gaussian states
\be
\psi(x|q_0,p_0,a,b)\equiv \frac{1}{(\pi a)^{1/4}}e^{\frac{-(x-q_0)^2}{2 a}+i p_0 (x-q_0)+ i b 
(x-q_0)^2 },
\ee
whose Wigner distribution is the two-dimensional Gaussian function
\be
W(q,p|q_0,p_0,a,b)=\frac{1}{\pi}e^{-\frac{(q-q_0)^2}a-a[p-p_0-2 b(q-q_0)]^2}.
\ee
$q_0$ and $p_0$ are the mean values of $q$ and $p$, $a$ and $b$ set the squeezing and the symmetry
axes of the distribution. 

From Eq.~(\ref{Wigner2}) it easy to verify that
\bey
\label{margi_x}
\int_{-\infty}^\infty dp W(q,p)=|\psi(q)|^2, \\
\int_{-\infty}^\infty dq W(q,p)=|\tilde\psi(p)|^2, 
\eey
where $\tilde\psi$ is the Fourier transform of $\psi$. Thus, the marginal probability distributions 
of $q$ and $p$ are the probability distributions of the observables $\hat x$ and $\hat p$,
respectively. In general, the probability distribution of the observables
$\cos\theta \hat q+\sin\theta\hat p$, called in 
quantum optics {\it quadratures}, is the marginal probability distribution of 
$\cos\theta x+\sin\theta p$. 

If only quadrature measurements of are considered, then the probability distribution
of the outgoing values can be obtained by means of the classical probability rules of 
Section~\ref{subsect_measure}. For example, we have from Eq.~(\ref{margi_x})
\be
|\psi(q)|^2=\int d\bar q d\bar p P(q|\bar q,\bar p)W(\bar q,\bar p),
\ee
where $P(q|\bar q,\bar p)\equiv \delta(q-\bar q)$ is the conditional probability density associated
with the measurement of $\hat q$.
Furthermore, it is possible to prove that for any unitary evolution whose generator is quadratic
in $\hat a$ and $\hat a^\dagger$, the Wigner function evolves according to the rules of
Section~\ref{subsect_dynamics}. More precisely, the evolution equation of $W$ is a Liouville
equation~\cite{montina_dice}. 

The manifold of the Gaussian quantum states is four-dimensional,
conversely the number of classical variables is two. The dimensional reduction
becomes more drastic when $M$ modes are considered. In this case, the number of classical
variables grows linearly with $M$, conversely the dimension of the manifold of 
Gaussian quantum states grows quadratically. Multidimensional Wigner functions
are used for example in the study of Bose-Einstein condensates~\cite{steel,monti3}.

\section{Theorem on the dimension of the ontological space}

\subsection{Definitions and properties}
\label{sect_proof}
In Section~\ref{sect_class} we have established three general conditions of 
ontological theories of quantum mechanics, let us enumerate them.

\begin{enumerate}
\item Let $X$, $\cal H$ and $\cal C$  be a set of discrete and/or continuous variables, a $N$-dimensional 
Hilbert space and a set which contains at least one element.
There exists a functional
$|\psi\rangle\rightarrow\{\rho(X|\psi,\eta),\eta\in C\}$ which associates each quantum state 
$|\psi\rangle\in\cal H$ with a set of probability distributions of the variables $X$.

\item Let $\cal E$ be a set with at least one element. There exists a functional
$\hat U\rightarrow\{P(X|\bar X,\hat U,\chi),\chi\in{\cal E}\}$ which associates
each unitary operator $\hat U$ of $\cal H$ with a set of conditional probabilities
$P(X|\bar X,\hat U,\chi)$. These distributions satisfy the properties in 
Sect.~\ref{subsect_dynamics}.

\item Let $\hat P(\phi)$ be the trace-one projector $|\phi\rangle\langle\phi|$. There exists
a conditional probability $P_M(\phi|X)$ such that
\be
\int dX P_M(\phi|X)\rho(X|\psi,\eta)=\langle\psi|\hat P(\phi)|\psi\rangle=|\langle\phi|\psi\rangle|^2.
\ee
for every $\eta$ and $|\psi\rangle$.
\end{enumerate}

It is useful to introduce the following definitions:

{\bf Definition} $\bf 2$. We call ${\cal S}(X)$ the set of quantum states $|\psi\rangle$ such that 
$X\in I(\psi)$.

{\bf Definition} $\bf 3$. $\hat U{\cal S}(X)$ is the set of the vectors $\hat U|\psi\rangle$,
with $|\psi\rangle\in{\cal S}(X)$. In simple words, $\hat U{\cal S}(X)$ is the unitary evolution
of the set ${\cal S}(X)$.

By means of Properties $1$-$2$, it is trivial to prove

{\bf Property} $\bf 4$. If $P(X|\bar X,\hat U,\chi)\ne0$, then $\hat U{\cal S}(\bar X)={\cal S}(X)$.

Proof. It is sufficient to prove that a state $|\psi\rangle$ is in ${\cal S}(\bar X)$ if
and only if $\hat U|\psi\rangle\in{\cal S}(X)$. If $|\psi\rangle\in{\cal S}(\bar X)$, then
$\bar X\in I(\psi)$. By Property 1, $X\in I(\hat U\psi)$, i.e., $\hat U|\psi\rangle\in {\cal S}(X)$.
Similarly, it is proved by means of Property 2 that $\hat U|\psi\rangle\in{\cal S}(X)$ $\Rightarrow$
$|\psi\rangle\in{\cal S}(\bar X)$.

\begin{figure}[h!]
\epsfig{figure=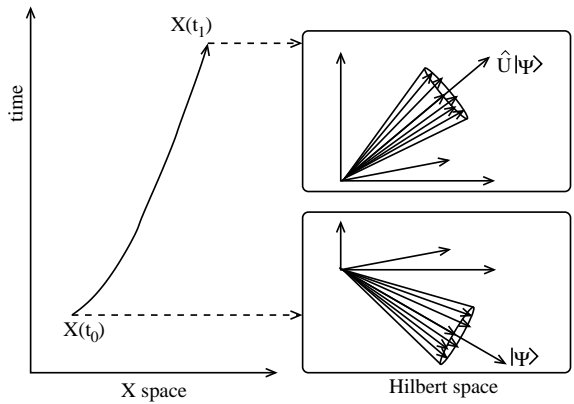,width=7.5cm}
\caption{Visual representation of a trajectory in a one-dimensional classical space and 
corresponding evolution of ${\cal S}$. At time $t_0$, the ontological variable takes the 
value $X(t_0)$. At
the subsequent time $t_1$, the variable has evolved to a value $X(t_1)$ with
probability $P[X(t_1)|X(t_0),\hat U,\chi]$. If this probability is finite, the
set ${\cal S}[X(t_1)]$ of Hilbert space vectors (top right inset) is equal to
the unitary evolution of ${\cal S}[X(t_0)]$ (bottom right inset), i.e.,
equal to $\hat U{\cal S}[X(t_0)]$. The sets ${\cal S}$  are represented as
cones with $|\Psi\rangle$ and $\hat U|\Psi\rangle$ as symmetry axis. Note that
if the Hilbert space has a dimension higher than $2$, the points of the
one-dimensional classical space can not map every possible orientation of
${\cal S}[X(t_1)]$.}
\label{fig2}
\end{figure}

Property 4 can be formulated in these terms. If $X(t)$ is a deterministic/stochastic trajectory in the 
ontic state space
and $\hat U(t_1;t_0)$ is the associated unitary evolution from time $t_0$ and to time $t_1$,
then
\be\label{unit_evo}
\hat U(t_1;t_0){\cal S}[X(t_0)]={\cal S}[X(t_1)],
\ee
for any $t_{0,1}$. An illustration of this property is reported in Fig. \ref{fig2}, where
the unitary evolution $\hat U(t_1;t_0)$ is synthetically denoted by $\hat U$.
At left we have represented a trajectory in a one-dimensional classical space. At right
the corresponding evolution of ${\cal S}$ is sketched. ${\cal S}[X(t_{0,1})]$ are drawn as 
cones with $|\Psi\rangle$ and $\hat U|\Psi\rangle$ as symmetry axis. The Hilbert space is represented 
as a three-dimensional Euclidean space. 

Another fundamental property of ${\cal S}$ holds.

{\bf Property 5}. The set ${\cal S}(X)$ can not contain every vector of the Hilbert space.
This is a consequence of Property~$3$. If $|\psi\rangle$ is a vector of ${\cal S}(X)$, i.e.
$X\in I(\psi)$, then
any orthogonal vector of $|\psi\rangle$ is not an element of this set.

At this point, the proof of our theorem on the classical space dimension is very simple.
\subsection{The theorem}
By means of the outlined properties, we will prove the following

\textbf{Theorem:}  The number of continuous variables in the set $X$ can not be smaller 
than $2 N-2$.

In the proof, we will use Properties~4 and 5, which are a synthesis of
Properties~1-3. 

\textbf{Proof:}
Let $\bar X$ be a fixed value such that ${\cal S}(\bar X)$ contains at least one vector of the Hilbert 
space.
Property~4 says that for every unitary evolution $\hat U$ there exists a value $X$ such that
\be\label{para_rota}
\hat U{\cal S}(\bar X)={\cal S}(X),
\ee
This implies that the number of continuous ontological variables in $X$ ($x_1,...x_w$;
see Sec.~\ref{kine}) is at least
equal to the number of parameters required to specify the orientation of any unitary evolution
of ${\cal S}(\bar X)$ (see Fig.~\ref{fig2}). If ${\cal S}(\bar X)$ would be invariant with respect to 
every $\hat U$, one would find
$0$ as lowest bound, but this is not our case, since ${\cal S}(\bar X)$ is not a null
set and does not contains every vector (Property~5).
As discussed in the following, the lowest number of required orientation parameters
is $2 N-2$. Thus, the theorem is proved.  \ $\square$

In an $N$-dimensional Hilbert space the orientation of $\hat U{\cal S}(X)$ can be specified by $N-1$
orthogonal vectors, which evolve according to $\hat U$, i.e., they are rigidly fixed in 
$\hat U{\cal S}(X)$.
This representation is sufficient, but could be redundant when ${\cal S}$ has some symmetry. In
order to clarify intuitively this point, one can consider some visual examples in the Euclidean
three-dimensional space. The orientation of a sphere does not require any parameter, since it is invariant
with respect to every rotation. A cylinder or a cone are invariant with respect to the rotation
around their symmetry axis. Thus, in order to specify their orientation it is sufficient to give
the direction of this axis, i.e., a vector. By contrast, a pyramid has no rotational symmetry
and we have to rigidly fix two orthogonal axes and specify their directions. In general, for a set of
elements in an $N$-dimensional Euclidean or Hilbert space the orientation is specified by
$N-1$ orthonormal vectors when the set has no symmetry with respect to rotations, whereas no
parameter is required if the set is completely symmetric. In our case, since ${\cal S}$ is not
uniform, the number of required orientation vectors is at least one. This minimal labeling is
possible when there exists one vector $|\Psi\rangle$ such that ${\cal S}$ is invariant
with respect to every unitary evolution which leaves $|\Psi\rangle$ unchanged.
In simple words, the orientation can be merely
specified by one vector when there exists a symmetry "axis" and $|\Psi\rangle$ is the
direction of this "axis" (see Fig.~\ref{fig2}). Thus, the set $\hat U{\cal S}(\bar X)={\cal S}(X)$ 
is identified by $\hat U|\Psi\rangle$, where $|\Psi\rangle$ is the symmetry axis of ${\cal S}(\bar X)$.
This vector is defined by $N$ complex numbers, 
but because of the normalization of the quantum states and the irrelevance of their global phase, 
it can be labeled by 
$2 N-2$ real parameters, which are the minimal number of parameters required to specify
the orientation of $\hat U{\cal S}$.

Let us consider the two-state model of Section~\ref{sect2} as a practical illustration
of the theorem. The space
$X$ is the set of three-dimensional unit vectors $\vec v$ and each quantum state is associated
with the probability distribution in Eq.~(\ref{prob_2states}). In this case, the elements of
${\cal S}(\vec v)$ are the Hilbert space vectors $|\psi\rangle$ such that 
$|\langle\psi|\phi\rangle|^2>B\equiv 1/2$,
$|\phi\rangle$ being the state whose bloch vector, defined by Eq.~(\ref{bloch_vector}),
is $\vec v$. It is clear that $|\phi\rangle$ is the symmetry axis of ${\cal S}(\vec v)$ and
determines the orientation of the set. The dynamics of ${\cal S}(\vec v)$ is a mere rotation
and the axis evolves according with Eq.~(\ref{para_rota}). This model is
dispersion-free. For more general models, it is possible to have
different symmetric sets ${\cal S}(\vec v)$ with $0\le B<1/2$. However, $B$ can not be larger
than $1/2$, because of Property~3. For $B=0$, set ${\cal S}(\vec v)$ contains only the
vector $|\phi\rangle$ and the conditional probability defined by Eq.~(\ref{cond_prob_spin}) is
replaced by  $P(\phi|\vec v)\equiv(1/2)[1+\vec w(\phi)\cdot\vec v]$. It is possible to
have theories with an asymmetric set ${\cal S}$, but in this case the two dimensions of
the Bloch sphere are not sufficient. 

The hidden-variable model discussed by Aaronson in Ref.~\cite{aaronson} is another practical
illustration of our result. Let $|1\rangle$,...,$|N\rangle$ be a complete orthonormal
basis of a $N$-dimensional Hilbert space. In Ref.~\cite{aaronson}, the quantum state
$|\psi\rangle\equiv\alpha_1|1\rangle+...+\alpha_N |N\rangle$ is mapped to
a probability distribution $\rho(n)\equiv\alpha_n$ whose domain is a space of 
$N$ ontic states. The unitary evolution is mapped to a $N\times N$
stochastic matrix, which corresponds to the conditional probability $P$ introduced
in Section~\ref{subsect_dynamics}. As previous proven by Hardy~\cite{hardy}, this 
mapping of quantum state does not work, since the number of ontic states has to be 
infinite.
Furthermore, our theorem says that we need at least $2N-2$ continuous ontological
variables. Aaronson makes up for such lack of a suitable number of ontic states 
by assuming that the stochastic matrix has to depend on the
quantum state $|\psi\rangle$ (See page $4$ in Ref.~\cite{aaronson}). This corresponds
to assume that $|\psi\rangle$ is an ontological variable, as in the Bohm theory.
These models are typical examples of pilot-wave theories and
the dimension of their ontic state space is consistent with our constraint.

Every known classical formulation of quantum mechanics
satisfies our constraint on the dimension of the classical space. 
In section~\ref{sect_examples} and here, we have considered some examples, 
other examples are provided in Refs.~\cite{strocchi,heslot,jones,rowe},
where a $N$-dimensional Hilbert space is reduced to a classical phase space of 
$2N$ real variables. Our result is relevant also for quantum Monte Carlo methods. By means
of them, one tries to map a quantum dynamics to a classical one with a reduced phase space
dimension. We have proved that this mapping is not possible, unless some condition on the
probability distributions and the conditional probability distributions is discarded.
Quantum Monte Carlo methods in a reduced sampling space are introduced for
example in Ref.~\cite{nummet}. In these methods, the kinematics and dynamics conditions
in Sec.~\ref{sect_class} are fulfilled, but they do not provide positive conditional
probability distributions for measurements. As a consequence, they
are subject to the celebrated "sign problem" and have
a computational complexity which grows exponentially with the evolution
time and the number of particles. This complexity is not due to the
dimension of the sampling space, but to the necessity of an exponentially
large number of realizations in order to reduce the statistical errors.

\section{conclusions}
The proved theorem clearly shows that, if quantum mechanics is formulated as a Markovian realistic theory,
the corresponding phase space grows exponentially with the physical size of the system. This occurs
for theories which are able to describe every attainable unitary evolution.
The explosion of the variable number implies that an exact Monte Carlo approach, which simulates quantum
processes by means of realistic Markovian chains, is in general subject to an exponential
growth of numerical resources and integration time~\cite{note2}.

In some practical cases, polynomial
algorithms could be feasible, for example when the trajectory of the quantum state is not dense
in the Hilbert space or different states are indistinguishable experimentally. This is the case
for example with dilute Bose-Einstein condensates in the mean field approximation.
Decoherence could
actually play an important role in the complexity reduction for concrete systems. Roughly speaking,
decoherence is due to our inability to distinguish a pure quantum state from a mixture of
other quantum states~\cite{decohe}. When a system is composed of many particles, such as a macroscopic gas, 
this inability is not merely technical, but fundamental. Thus, a large set of states could be
described as statistical mixtures of a smaller set of pure states, enabling to simulate the dynamics
by means of a Markovian theory in a reduced phase space, as expected in classical limit.
However, it is generally surmised that 
systems as quantum computers require nearly unitary evolutions with negligible decoherence
effects in order to execute efficiently quantum algorithms~\cite{nielsen}
and would not be suitably simulated by approximated stochastic methods which replace pure state
by mixtures. These conclusions would support the conjecture that, in general, quantum algorithms cannot
be efficiently simulated by classical computers in polynomial time~\cite{bennett}.
Indeed, the actual speed-up of quantum computation requires further corroborations. 
For example, nobody has proven that factoring does not have a polynomial solution classically.
Note that the subset of available quantum states can be parametrized with a number of variables which
grows with the number of quantum gates, i.e., with the physical size.
If this subset is dense in the Hilbert space or its parametrization
is computational hard, then our theorem supports the conjecture on the polynomial non-computability,
but these conditions cannot be assumed {\it a priori}. The subset parametrization is certainly 
possible for example in particular classes of quantum computers, as recently reported 
in Refs.~\cite{vidal,yoran}.

Finally, we conclude with a discussion on some possible implications of our theorem 
for the future study of the hidden variable theories. As said in the introduction, 
the main criticism against the known ontological models, such as Bohm mechanics, is 
indeed the exponential growth of the ontic space dimension with the physical size.
We have shown rigorously that this feature is unavoidable in the framework of 
causal Markovian theories. Thus our result seems to put another nail in the hidden
variable coffin and to imply that the realistic interpretations do not provide
a practical advantage for the study of quantum systems. However, our intent
is to give a constructive result and to suggest a novel direction for the 
investigations on the ontological theories. 
In order to avoid the exponential growth of the number of ontic variables, we have
one possibility, to discard some hypotheses of the theorem. In our opinion, the 
Markovian property is the only one sacrifiable. More drastically, we could discard
the causality hypothesis. It is interesting to observe that the Bell theorem
and the Lorentz invariance seem to suggest the same conclusion. The Bell theorem
establishes that an ontological theory of quantum mechanics can not be local
and relativity implies that a non-local theory is also non-causal. 
We suspect that realism in quantum mechanics with non-pathological consequences
for the ontic space dimension could be possible with non-causal rules for 
evaluating correlations among events.

I thank F. T. Arecchi for useful discussions. This work was supported by Ente Cassa di Risparmio di
Firenze under the project "dinamiche cerebrali caotiche".


\begin{thebibliography}{20}
\bibitem{note2} The exponential growth of the integration time is implied by the necessity of reading the
whole information on $X$ in order to evaluate the variables at the next time.
\end{thebibliography}
\end{document}